\documentclass[12pt]{article}
\usepackage{amsmath,amssymb}
\usepackage{graphicx,psfrag,epsf}
\usepackage{enumerate}
\usepackage{mathtools,tikz-cd}
\usepackage{algorithm}
\usepackage{algorithmic}
\usepackage{subcaption}
\usepackage{authblk}
\usepackage{url} 

\newcommand{\blind}{1}

\newcommand{\argmax}{\arg\!\max}
\newtheorem{definition}{Definition}
\begin{document}

\def\spacingset#1{\renewcommand{\baselinestretch}%
{#1}\small\normalsize} \spacingset{1}



\if1\blind
{
  \title{\bf Divide and Recombine for Large and Complex Data: Model Likelihood Functions using MCMC}
  \author[1]{Qi Liu}
  \author[1]{Anindya Bhadra}
   \author[1]{William S. Cleveland}
   \affil[1]{Department of Statistics, Purdue University}
  \maketitle
} \fi

\if0\blind
{
  \bigskip
  \bigskip
  \bigskip
  \begin{center}
    {\LARGE\bf Title}
\end{center}
  \medskip
} \fi

\bigskip
\begin{abstract}
In Divide \& Recombine (D\&R), big data are divided into subsets, each analytic method is applied to subsets, and the outputs are recombined. This enables deep analysis and practical computational performance. An innovate D\&R procedure is proposed to compute likelihood functions of data-model (DM) parameters for big data. The likelihood-model (LM) is a parametric probability density function of the DM parameters. The density parameters are estimated by fitting the density to MCMC
draws from each subset DM likelihood function, and then the fitted densities are recombined. The procedure is illustrated using normal and skew-normal LMs for the logistic regression DM.
\end{abstract}

\noindent%
\vfill

\newpage
\spacingset{1.45} 
\section{Introduction}
For big data, analysis creates immense computation challenges which can be computed too long that is impractical or even worse, infeasible. One example is computing likelihood function for both estimation and inference, which is now suffering as a result of the huge computational demand. Likelihood modeling within divide and recombine (D\&R) provides feasible, practical computation strategies to accelerate computation.

The fundamental idea for the likelihood modeling within D\&R framework is as follows.
Suppose that the data consist of N conditionally independent observations. Each observation contains explanatory variables $x_i \in \mathbb{R}^p$ (including intercept) and response variable $y_i$. The likelihood function for some parametric data model is a function of coefficient parameters $\theta$ given by
  $$ L(\theta) = \prod_{i=1}^N L(\theta |x_i, y_i) $$
We assume that the dataset (X, Y) is too large to reside in a singe machine. Therefore, it is divided into R subsets: $(X_1,Y_1), \dots, (X_R,Y_R)$, each with M observations, such that $(x_{(s)i}, y_{(s)i})$ is the i-th observation of the subset $(X_s,Y_s)$. Thus, the all-data likelihood function is given by
\begin{equation}
L(\theta) = \prod_{s=1}^R L_{(s)}(\theta),
\end{equation}
which we refer to as the independent product equation, where $ L_{(s)}(\theta)$ is the subset likelihood function defined by
\begin{align*}
L_{(s)}(\theta)&= \prod_{i=1}^M L(\theta |x_{(s)i}, y_{(s)i}). 
\end{align*}
This equation indicates that under the independence assumption, the likelihood of the full data can be represented by the product of subset likelihood functions. In likelihood modeling, we work with some parameterized class of distributions $g(\theta | \phi)$, where $\phi$ is the parameter of density function (e.g. mean and covariance matrix in the Gaussian density function). For each subset, the density parameters for pre-chosen density family are estimated by fitting the density to MCMC draws from each subset DM likelihood function. Then 
$$g_{(s)}(\theta | \hat{\phi)} \approx  C_{s}\times L_{(s)}(\theta).$$
Finally, the full-data likelihood function can be approximated by the product of the subset fitted density functions, up to a multiplicative constant. 
\begin{equation}
L(\theta) \approx  \prod_{s=1}^R \frac{1}{C_{s}} g_{(s)}(\theta | \hat{\phi)} = C\times \prod_{s=1}^R g_{(s)}(\theta | \hat{\phi)}.
\end{equation}
There are many candidate distributions $g(\theta | \phi)$, just as there are many models for DM. Of course, one thing is attempting to try is normal density as the likelihood function tends to normal when n becomes big. There are two fundamental questions:
\begin{enumerate}
\item How to assess whether some candidate distribution well approximates the subset likelihood function? 
\item How close to the full-data likelihood function the approximated recombined likelihood function is?
\end{enumerate}
 To answer these two questions, we propose the contour probability algorithm to visually quantify the distance between two unnormalized density functions. The model diagnostics are applied to both subset likelihood modeling and the final all-data likelihood modeling.

The remainder of this article is organized as follows. In Section 2, normal and skew-normal are presented to illustrate the choice of LM. Section 3 addresses how to merge approximate subset likelihoods to formulate an approximate all-data likelihood. And the likelihood modeling algorithm is proposed for the skew-normal family. LM diagnostic method -- contour probability algorithm is discussed in detail in section 4. Section 5 provides a real data example illustrating that the skew-normal likelihood modeling better captures the posterior density, as well as a variety of simulated datasets to assess the performance of the likelihood modeling. Section 6 is a concluding discussion.
 \label{sec:intro}

\section{The Choice of LM}
Model building is used for LM, including diagnostic methods to check how well
LM fits the subset likelihoods and full-data likelihood. This is just like model building and checking
for the DM, although the details for the diagnostics are not the same.

There are many candidates, just as there are many models for DM. Normal and skew-normal are presented here as illustrations. The modeling building and checking can, as with a DM, lead to insight about a better LM.

\subsection{Normal Family}
One thing is attempting to try is normal density as the likelihood function
tends to normal when n becomes big. Our objective is to find
$$N(\theta |\mu, \Sigma) \rightarrow L(\theta |X_s, Y_s)$$
where $\mu$ and $\Sigma$ are the mean and covariance matrix of the normal distribution. 

There are two approaches to estimate the parameters in the normal density function. One is to match the mode of the normal density to the mode for the subset likelihood function, which is computed by maximum likelihood estimation (MLE); and estimate the covariance matrix as a function of the Hessian matrix evaluated at the MLE. We refer this method as \textbf{Local Information} (Local) method. This method is equivalent to
approximate the subset likelihood function by using a normal density with a mean (the subset MLE), and
variance matrix (inverse of the observed Fisher information), up to a constant multiplier.
\begin{align*}
   \hat{\mu} &=  \argmax_{\theta} l(\theta |X_r, Y_r)\\
   \hat{\Sigma} &= \mathcal{I}^{-1}
   \end{align*}
where $\mathcal{I}$ is the observed Fisher information. Another approach is to generate a sample according to the stationary function $L(\theta |X_s, Y_s)$ using Markov chain Monte Carlo (MCMC) method, and estimate $(\hat{\mu}, \hat{\Sigma})$ using the sample moments. We call it \textbf{Moment Matching} (MM) method.

The inference based on the normality might be not reliable if the departure from the normal assumption of the subset likelihood is serious because the model can be very complex and the subset data based on some divisions might be not large enough. Therefore, we propose a more general density family -- Skew-normal family to model likelihoods.

\subsection{Skew-normal (SN) Family}
Generally, the method of moments (MM) and the MLE (Local) are two widely used methods for estimation of population density parameters. The MM is preferable for the skew-normal family due to following reasons. For statistical inference, one concerns the behavior of the likelihood function and other related quantities for a sample from the SN distribution in the neighborhood of $\alpha = 0$  (the shape parameter in the skew-normal density function), a value of particular relevance since there the SN family reduces to the normal one. First, a sort of non-quadratic shape of the log-likelihood function has been exhibited with many data in Azzalini et al. (\cite{Azzalini.2008} 2008). Another unpleasant phenomenon is that, at $\alpha = 0$, the expected Fisher information is singular, even if all parameters are identifiable. Moreover, closed-form solutions for the maximum likelihood estimator do not exist. Therefore, we estimate parameters of the skew-normal using the MM method instead of the Local method. 

The multivariate SN distribution has been widely discussed by Azzalini, Dalla Valle and Capitanio (\cite{Azzalini.1996} 1996; \cite{Azzalini.1999} 1999). The p-dimensional SN density function is defined by
$$
f_p(\theta|\xi,\Omega,\alpha) =\frac {2}{\sqrt{(2 \pi)^p |\Omega|}} \exp \left( -\frac{1}{2} (\theta - \xi)^\intercal \Omega^{-1} (\theta - \xi) \right)\Phi(\alpha^\intercal \omega^{-1}(\theta -\xi)), \, \xi, \alpha \in \mathbb{R}^p, \Omega \in \mathbb{R}^{p\times p},
$$
where $\Omega$ is a $p\times p$ positive definite matrix, $\xi$ is a vector location parameter, $ \alpha$ is a vector shape parameter, and $\omega$ is a diagonal matrix formed by the square root of the diagonal of $\Omega$. We say $\Theta \sim SN(\xi,\Omega, \alpha)$ if a multivariate random variable $\Theta$ has density function $f_p(\theta|\xi,\Omega,\alpha)$.

Given a sample generated from $L(\theta |X_s, Y_s)$ using Markov chain Monte Carlo (MCMC) method, sample mean $\hat{\mu}_\Theta$, sample covariance $\hat{\Sigma}_\Theta$, and component-wise skewness $\hat{\gamma}_{\Theta}$ can be easily computed. There is a mapping:
$$ (\hat{\xi}, \hat{\Omega}, \hat{\alpha}) \rightarrow (\hat{\mu}_\Theta, \hat{\Sigma}_\Theta, \hat{\gamma}_{\Theta}).$$
However, not vice versa. In order to obtain the parameters estimates, we resample the data until $(\hat{\xi}, \hat{\Omega}, \hat{\alpha})$ can be estimated. The detail derivations for the parameter estimation for the Skew-normal is illustrated in Appendix.

 \section{Recombination}
\begin{figure}[h]\label{full_diagram}
\centering
\includegraphics[scale = 0.4]{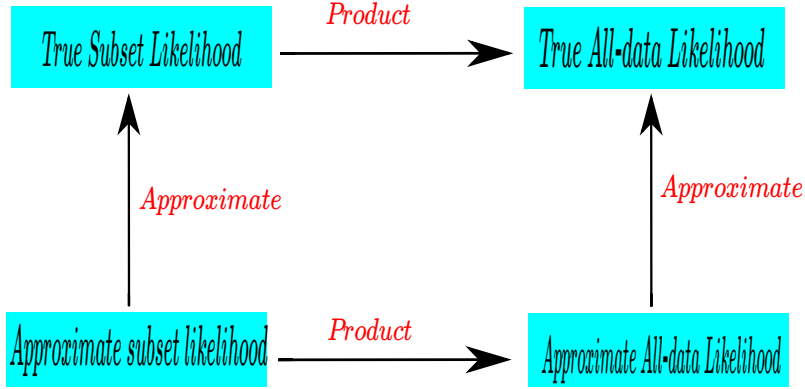}
\caption{A diagram of likelihood modeling for big data}
\end{figure}

 In this section, we will address how to merge approximate subset likelihoods to formulate an approximate all-data likelihood function such that the overall quality of inference is good comparing the one for the true likelihood function. The subset likelihood is, in general, a nontrivial function of all of the data in
a given subset. It can not be expressed without reading all of the data. Therfore, the subset likelihood modelling is introduced to model each subset likelihood on some distribution family such that each fitted
subset likelihood can be expressed by only a small number of distribution parameters, up to a multiplicative constant (left bottom to left top in Figure \ref{full_diagram}). The approximation of full-data likelihood is the product of approximate subset likelihoods (right bottom to right top in Figure \ref{full_diagram}). We will investigate two likelihood models in detail: skew-normal model and normal model.
\subsection{Normal Moment Matching Estimation}
 Recall that the likelihood function for each subset is given by
 $$
 L_{(s)}(\theta)= \prod_{i=1}^M L(\theta |x_{(s)i}, y_{(s)i}). 
 $$
 which is a function of $\theta$. Assume that subset likelihood function is approximated by the normal density function, up to a multiplicative constant. The all-data likelihood function is approximated by
 $$
 L^{Norm}(\theta) = \prod_{s=1}^R N(\theta|\hat{\mu}, \hat{\Sigma}),
 $$
Which is also normal density function, up to a multiplicative constant. Therefore, the recombined approximate log likelihood for the normal model is 
$$ l^{Norm}(\theta) = \log{L^{Norm}(\theta)}= c -\frac{1}{2} (\theta - \hat{\mu})^\intercal \hat{\Sigma}^{-1} (\theta - \hat{\mu}),$$
where $$\hat{\Sigma}^{-1} = \sum_{s=1}^R \hat{\Sigma}^{-1}_{(s)}, \quad \hat{\mu} = \hat{\Sigma} \sum_{s=1}^R\hat{\Sigma}^{-1}_{(s)} \hat{\mu}_{(s)}.$$  
$(\hat{\mu}_{(s)}, \hat{\Sigma}_{(s)})$ is estimated by using sample mean and sample covariance matrix of the MCMC sampling of the subset likelihood function; c is a constant.
\begin{definition}
The normal $D\&R$ estimate using the MM method (NMM) is defined by
$$\hat{\theta}_\text{NMM} = \arg \max_\theta l^{Norm}(\theta) = \hat{\mu}.$$
\end{definition}
 \subsection{Skew-normal Moment Matching Estimation}
 Assume that the subset likelihood model is the skew-normal model,
then $L_{(s)}(\theta)$ is approximated by the skew-normal $SN(\theta|\hat{\xi}_{(s)}, \hat{\Omega}_{(s)}, \hat{\alpha}_{(s)})$, up to a multiplicative constant. Therefore, the all-data likelihood function is approximated by
 $$
 L^{SN}(\theta) = \prod_{s=1}^R SN(\theta|\hat{\xi}_{(s)}, \hat{\Omega}_{(s)}, \hat{\alpha}_{(s)}).
 $$
The recombined approximate log likelihood for the skew-normal model is
\begin{equation}\label{sn.loglike}
 l^{SN}(\theta) = \sum_{s=1}^R \log SN(\theta|\hat{\xi}_{(s)}, \hat{\Omega}_{(s)}, \hat{\alpha}_{(s)}) = c -\frac{1}{2} (\theta - \hat{\xi})^\intercal \hat{\Omega}^{-1} (\theta - \hat{\xi})  +
  \sum_{s=1}^{R} \log \Phi \left(\hat{\lambda}_{(s)}^\intercal (\theta - \hat{\xi}_{(s)})\right), 
  \end{equation}
where
\begin{align*} 
\hat{\Omega}^{-1} &= \sum_{s=1}^R \hat{\Omega}^{-1}_{(s)},\\
 \hat{\lambda}_{(s)}^\intercal &= \hat{\alpha}_{(s)}^\intercal \hat{\omega}_{(s)}^{-1},\\
 \hat{\xi} &= \hat{\Omega} \sum_{s=1}^R \hat{\Omega}^{-1}_{(s)} \hat{\xi}_{(s)}. 
 \end{align*}
$(\hat{\xi}_{(s)}, \hat{\Omega}^{-1}_{(s)}, \hat{\alpha}_{(s)})$ is estimated by using formulas (13)-(15) in the Appendix if $p = 1$ or (16)-(18) if $p > 1$; c is a constant and $ \hat{\omega}_{(s)}$ is the diagonal matrix formed by the square root of the diagonal of $\hat{\Omega}_{(s)}$. 
\begin{definition}
The skew-normal $D\&R$ estimate using the MM method (SNMM) is defined by
\begin{equation}\label{snest}
\hat{\theta}_\text{SNMM} = \arg \max_\theta l^{SN}(\theta).
\end{equation}
\end{definition}

How do we know the SNMM is well defined? Actually, $l^{SN}(\theta)$ is a concave function because it is the sum of log skew normal density functions which are concave. Therefore, the recombined approximate log-likelihood for the skew-normal model is unimodal. The proof of the concavity of the multivariate SN density is provided in the Appendix.

From the general theory about the MLE, the sampling distribution of a MLE is approximately normal. And the asymptotic estimated covariance matrix for the coefficient parameter estimates is obtained from the Fisher scoring estimation method. Specifically, the asymptotic covariance matrix is given by a function of the information matrix. Based on above approximate log likelihood function, the observed Fisher information matrix can be estimated by 
\begin{equation*}
\mathcal{I} = - \frac{\partial^2}{\partial \theta \partial \theta^T }l^{SN}(\theta)= \hat{\Omega}^{-1} - \sum_{s=1}^{R}\frac { \phi_{(s)}'(\hat{\lambda}_{(s)}^\intercal (\theta - \hat{\xi}_{(s)})) \Phi_{(s)}(\hat{\lambda}_{(s)}^\intercal (\theta - \hat{\xi}_{(s)})) - \phi_{(s)}^2(\hat{\lambda}_{(s)}^\intercal (\theta - \hat{\xi}_{(s)}))} {\Phi_{(s)}^2(\hat{\lambda}_{(s)}^\intercal (\theta - \hat{\xi}_{(s)})) }\hat{\lambda}_{(s)}\hat{\lambda}_{(s)}^\intercal,
\end{equation*}
where 
$$ \phi_{(s)}(\hat{\lambda}_{(s)}^\intercal (\theta - \hat{\xi}_{(s)})) = \frac {1} {\sqrt{2 \pi}} e^{ - \frac{1}{2} (\hat{\lambda}_{(s)}^\intercal (\theta - \hat{\xi}_{(s)}))^2 }, $$

$$ \Phi_{(s)}(\hat{\lambda}_{(s)}^\intercal (\theta - \hat{\xi}_{(s)})) = \int_{- \infty}^{ \hat{\lambda}_{(s)}^\intercal (\theta - \hat{\xi}_{(s)})} \frac {1} {\sqrt{2 \pi}} e^{ - \frac{1}{2} x^2 } dx, $$

$$ \phi_{(s)}'(\hat{\lambda}_{(s)}^\intercal (\theta - \hat{\xi}_{(s)})) = \frac {-1} {\sqrt{2 \pi}} e^{ - \frac{1}{2} (\hat{\lambda}_{(s)}^\intercal (\theta - \hat{\xi}_{(s)}))^2 } \hat{\lambda}_{(s)}^\intercal (\theta - \hat{\xi}_{(s)})).$$
Therefore, 
\begin{equation}\label{cov}
\hat{\theta}_\text{SNMM} \longrightarrow^{L} N(\theta,\mathcal{I}^{-1}).
\end{equation}

In real world applications, the optimizer of (\ref{sn.loglike}) is not easy to compute when the number of subsets R is large. For this scenario, we propose a simplified version of the recombined log likelihood for the skew-normal model as follows:
\begin{equation*}
 l^{SSN}(\theta) =c -\frac{1}{2} (\theta - \hat{\xi})^\intercal \hat{\Omega}^{-1} (\theta - \hat{\xi})  +
   R\times\log \Phi \left(\hat{\lambda}_A^\intercal (\theta - \hat{\xi}_A)\right),
  \end{equation*}
where
\begin{align*} 
\hat{\lambda}_A^\intercal &=\frac{\sum_{s=1}^R \hat{\lambda}_{(s)}^\intercal}{R},  \\
\hat{\xi}_A &=  \sum_{s=1}^R  \hat{\xi}_{(s)} /R.
 \end{align*}
 
 \begin{definition}
The simplified skew-normal $D\&R$ estimate using the MM method (SSNMM) is defined by
 \begin{equation}
\hat{\theta}_\text{SSNMM} = \arg \max_\theta l^{SSN}(\theta).
\end{equation}
\end{definition}
 
From a Bayesian perspective, the likelihood function is proportional to the posterior density function when the prior is the uniform distribution. Therefore, the recombined likelihood function provides a good approximate posterior density function, which can be used to perform statistical inference such as posterior mean estimation, credible interval computation and hypothesis testing. 

\begin{algorithm}
\caption{Likelihood Model Fitting Procedure using Skew-normal Density }\label{alg:LM}
\begin{algorithmic}
\REQUIRE $X, Y$ \quad \COMMENT{$X \in \mathbb{R}^{N\times p}$ and $Y \in R^{N}$}
\STATE Divide $(X,Y)$ into R submatrix $X_i \in \mathbb{R}^{M_i\times p}, Y_i\in R^{M_i}, i = 1,\dots,R$
\STATE The following for loop is computed in parallel
\FOR{$s = 1\colon R$} 
\STATE Generate MCMC draws according to the stationary function $L_{(s)}(\theta)$
\STATE Estimate $(\hat{\xi}_{(s)}, \hat{\omega}_{(s)}, \hat{\alpha}_{(s)})$ using MCMC draws
\ENDFOR
\STATE Recombine subset approximate likelihoods to formulate the log of approximate likelihood $l^{SN}(\theta)$
\STATE Calculate the SNMM $\hat{\theta}_\text{SNMM}$ based on (\ref{snest}), and its covariance matrix $Cov(\hat{\theta}_\text{SNMM})$ using the observed Fisher information
\RETURN $(\hat{\theta}_\text{SNMM}, Cov(\hat{\theta}_\text{SNMM}))$\quad \COMMENT{Normal density function with the mean and the variance}
\end{algorithmic}
\end{algorithm}

\section{LM Diagnostics -- Contour Probability Algorithm}
For univariate likelihood functions, the visible comparison between approximate likelihood and true likelihood can be achieved by plotting log likelihood ratio over a neighborhood of the MLE. In contrast, it is a big challenge to visualize how close one likelihood function is to another likelihood function when the dimension of the parameter vector is high. In the case of one-dimensional distributions, the Kolmogorov-Smirnov (K-S) test by Massey 1951 \cite{Massey.1951}, is based on the maximum distance between the cumulative distribution functions of two histograms or probability densities. The K-S test is non-parametric and independent of the shapes of the underlying distributions. However, it does not generalize naturally to higher dimensions, and there is no widely accepted test for comparing N-dimensional distributions (Loudin et al., 2003 \cite{loudin2003}). Another popular method is the likelihood ratio test. However, for our case, it requires computing normalizing constant of the likelihood function, which is computationally intense and numerically unstable for high dimensional functions, such as the logistic likelihood function, with a huge number of observations.

A new method is proposed to measure the similarity between approximate multivariate likelihood function and the true multivariate likelihood function without calculating the corresponding normalizing constants. Instead of using the difference between the empirical distribution function of the sample of the approximate likelihood function and the cumulative distribution function of the true likelihood distribution, we consider a series of probabilities that samples drawn from the approximate likelihood fall in regions bounded by predefined high dimensional ellipsoids, respectively. What is the contour probability? Why can contour probabilities measure the difference between two likelihood functions? 
\begin{figure}[h]
\centering
\includegraphics[width=4in, angle=-90]{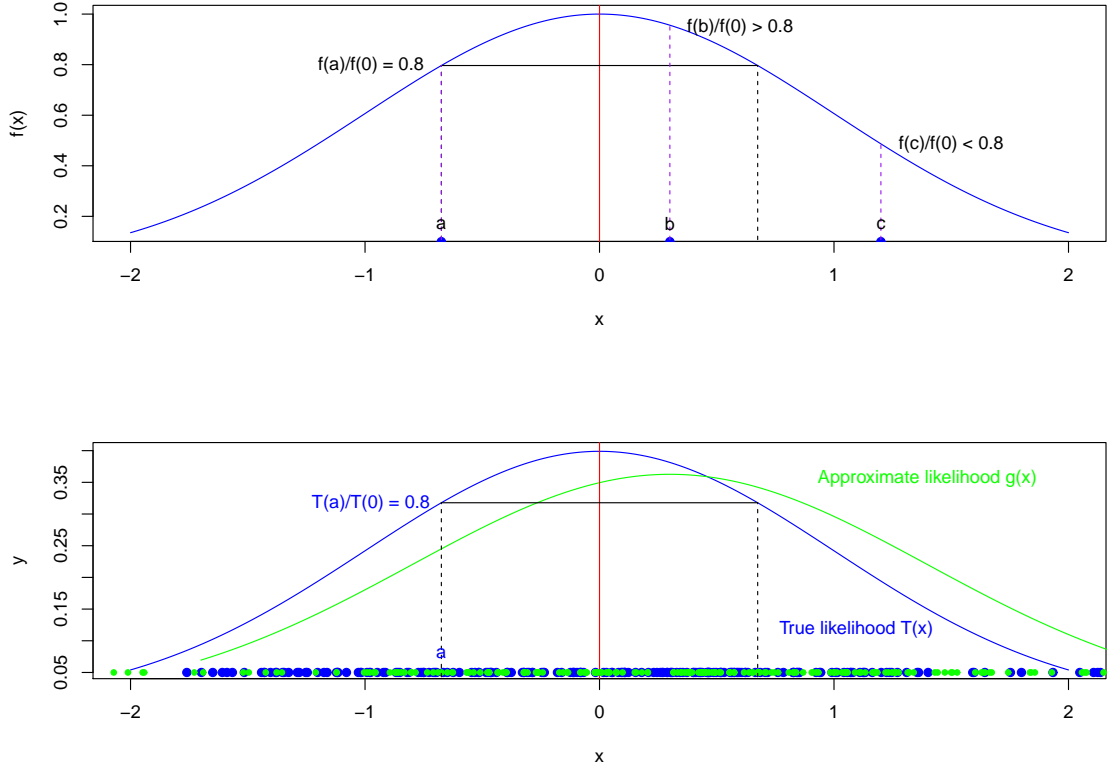}
\caption{The upper panel displays the plot for $f(x) = e^{-\frac{x^2}{2}}$. In the lower panel, T(x) is the reference density function, which is the standard normal density function, while g(x) is the approximate density function which is the normal density function with mean 0.3 and standard error 1.1. The blue dots on the bottom are a random sample generated from T(x) and the green ones are from g(x).}
\label{fig:contour.norm}
\end{figure}

The idea of the contour probability is motivated by the Monte Carlo method. Take a univariate normal density function as an example. In Figure \ref{fig:contour.norm}, the upper panel is a plot for the function $f(x)$. Suppose the normalizing constant C is unknown even though it is known to be $\sqrt{2\pi}$, how to calculate $E = \int_{-b}^{b}\frac{f(x)}{C}dx$? The principle of the Monte Carlo method \cite{Robert2005} for approximating $E$ is to generate a sample $(x_1, \cdots, x_n)$ from the $f(x)$ and propose the empirical average as an approximation
$$\hat{E} = \frac{\sum_{i=1}^n I_{\vert x_i\vert<\vert a\vert}}{n}.$$
As $f(x)$ is concave, it is equivalent to 
$$\hat{E} = \frac{\sum_{i=1}^n I_{f(x_i)/f(0) > 0.8}}{n} $$
where I is an indicator function. For a given ratio $h \in (0,1)$, $A_h = \{x \vert f(x)/f(0) > h\}$ is a region bounded by a contour, and there is only one corresponding probability $E_h = \int_{A_h}\frac{f(x)}{C}dx$. Therefore, there is a mapping 
$$CP: h \in (0, 1)\rightarrow E_h\in(0,1)$$
It is worth noting that the probability is estimated by using the sample generated from the target function, without knowing the normalizing constant. Also, this method can be naturally generalized to multivariate concave positive functions.


In order to demonstrate how the contour probabilities can measure the difference between two functions, we consider the probability density function of N(0,1) and N(0.3,11) as the reference function and the approximate function, respectively, which are displayed in the lower panel of Figure \ref{fig:contour.norm}. Assume a sample $(x_1, \cdots, x_n)$ and a sample $(y_1, \cdots, y_n)$ are drawn from $T(x)$ and $g(y)$, respectively. For a given $h = 0.8$, $A_h = \{x \vert T(x)/T(0) > h\} = (a, -a)$. Then $E_T = \int_{a}^{-a}T(x)dx$ and $E_g = \int_{a}^{-a}g(y)dy$ can be estimated by
\begin{align*}
\hat{E_T} &= \frac{\sum_{i=1}^n I_{\vert x_i\vert<\vert a\vert}}{n} \Longleftrightarrow \hat{E_T} = \frac{\sum_{i=1}^n I_{T(x_i)/T(0) > 0.8}}{n}\\
 \hat{E_g} &= \frac{\sum_{i=1}^n I_{\vert y_i\vert<\vert a\vert}}{n} \Longleftrightarrow \hat{E_g} = \frac{\sum_{i=1}^n I_{T(y_i)/T(0) > 0.8}}{n}
\end{align*}
Therefore, there will be a pair of probabilities $(\hat{E_T}(h), \hat{E_g}(h))$ for any given ratio $h \in (0, 1)$. A series of points $(\hat{E_T}(h), \hat{E_g}(h))$ are supposed to lie around the straight line $y = x$ in that $\hat{E_g}$ is supposed to be close to $\hat{E_T}$ if g(x) well approximates T(x). Alternatively, if the contour probability difference is plotted against the contour probability of T(x), i.e. $(\hat{E_g}(h)- \hat{E_T}(h), \hat{E_T}(h))$, the points should be not far away from $y = 0$.

All of above reasoning suggests the contour probability algorithm (CPA) in Algorithm \ref{alg:CPA}. $L(\theta)$ and $L^{approx}(\theta)$ are the true likelihood function and approximate likelihood function, respectively. Assume $L(\theta)$ is unimodal.

\begin{algorithm}
\caption{Contour Probability Algorithm (CPA) }\label{alg:CPA}
\begin{algorithmic}
\REQUIRE $h_i \in (0,1), i = 1, \cdots, k $, $L(\theta)$ and $L^{approx}(\theta)$
\STATE Draw a sample $(\theta_1, \cdots, \theta_{n1})$ and a sample $(\theta_1^a, \cdots, \theta_{n2}^a)$ from $L(\theta)$ and $L^{approx}(\theta)$, respectively
\STATE Compute MLE of $L(\theta)$ denoted by $\hat{\theta}_\text{MLE}$
\FOR{$i = 1\colon k$} 
\STATE Count the number of the points $\tilde{\theta}$ satisfying 
 $$\frac{L(\tilde{\theta})}{L(\hat{\theta}_\text{MLE})} > h_i  \Longleftrightarrow l(\tilde{\theta}) - l(\hat{\theta}_\text{MLE}) > log(h_i) $$
 in both the approximate likelihood sample and the true likelihood sample, denoted by $a_i$ and $t_i$, respectively.  
\STATE $A_i = \frac{a_i}{n_2}, T_i = \frac{t_i}{n_1}$
\ENDFOR
\RETURN $A = (A_1, \cdots, A_k), T = (T_1, \cdots, T_k), $
\end{algorithmic}
\end{algorithm}

\section{Real Data and Simulated Experiments}
This section proceeds through a real data example illustrating the contour probability algorithm and simulated examples for logistic regression to assess the performance of likelihood modeling on big data.
\subsection{Data and Model}
We use one simple example to show how skew-normal likelihood modeling can capture more information of subset likelihoods or subset posterior densities. The data are the summary of exit polls in 58 counties in California (see Appendix C). The polls were conducted several hours before the end of the primary on June 7, 2016, with the total number of sampled people in each county fixed by design. The goal is to predict Hillary Clinton’s vote share in each county, as well as her vote share in California overall. The data include following variables.

\begin{itemize}
\item Fips ($j$): The Federal Information Processing Standard (FIPS) code that uniquely identifies a county in the United States.

\item Total voters ($N_j$): The total number of registered voters in the California Democratic primary.

\item Sample voters ($n_j$): The total number of voters in the exit poll.

\item Sample clinton ($y_j$): The total number of votes for Clinton in the exit poll.
\end{itemize}

The data from counties $j = 1, . . . , J, J = 58$, are assumed to follow independent binomial distributions:
$$y_j | \theta_j \sim Binomial(n_j, \theta_j), \quad j = 1, \dots, 58,$$
with the number of sample votes, $n_j$, known. The parameters $\theta_j$ are assumed to be independent
samples from a beta distribution:
$$\theta_j | \alpha, \beta \sim Beta(\alpha, \beta),$$
and we shall assign a noninformative hyper-prior distribution to reflect our ignorance about
the unknown hyper-parameters. However, we must check that the posterior distribution is proper.
One reasonable choice of the hyper-prior density of $(\alpha, \beta)$ is
$$(\alpha, \beta) \sim (\alpha + \beta)^{-5/2}.$$
The corresponding posterior density is proper as long as $0 < y_j < n_j$ for at least one experiment $j$ \cite{Gelman2013}. Combining the sampling model for the observable $y_j'$s and the prior distribution yields
the joint posterior distribution of all the parameters and hyper-parameters, which can be expressed as follows  
\begin{align*}
p(\alpha, \beta, \theta_1, \dots, \theta_J) &\propto p(\alpha, \beta)\prod_{i=1}^J Binomial(y_i| \theta_i) Beta(\theta_i| \alpha, \beta) \\
&\propto (\alpha + \beta)^{-5/2} \prod_{i=1}^J{\frac{\Gamma(\alpha + \beta)}{\Gamma(\alpha) \Gamma(\beta)} \theta_i^{\alpha + y_i -1}(1-\theta_i)^{n_i+\beta - y_i -1}}.
\end{align*}
Thus we can write the marginal posterior density of the hyper-parameters as 
\begin{align}
p(\alpha, \beta | y)&\propto (\alpha + \beta)^{-5/2} \prod_{i=1}^J{\int \frac{\Gamma(\alpha + \beta)}{\Gamma(\alpha) \Gamma(\beta)} \theta_i^{\alpha + y_i -1}(1-\theta_i)^{n_i+\beta - y_i -1}d\theta_i}\\
&\propto (\alpha + \beta)^{-5/2}(\frac{\Gamma(\alpha + \beta)}{\Gamma(\alpha) \Gamma(\beta)})^J \prod_{i=1}^J\frac{\Gamma(\alpha + y_i)\Gamma(\beta + n_i-y_i)}{\Gamma(\alpha + \beta + n_i)}
\end{align}

\subsection{Approximate Methods for Posterior Distribution}
In this section, \textbf{Local Information}, \textbf{Moment Matching} methods with the normal family, and \textbf{Moment Matching} with the SN family are applied to approximate the posterior density. 
\begin{figure}[h]
\centering
\includegraphics[width=3in, angle=-90]{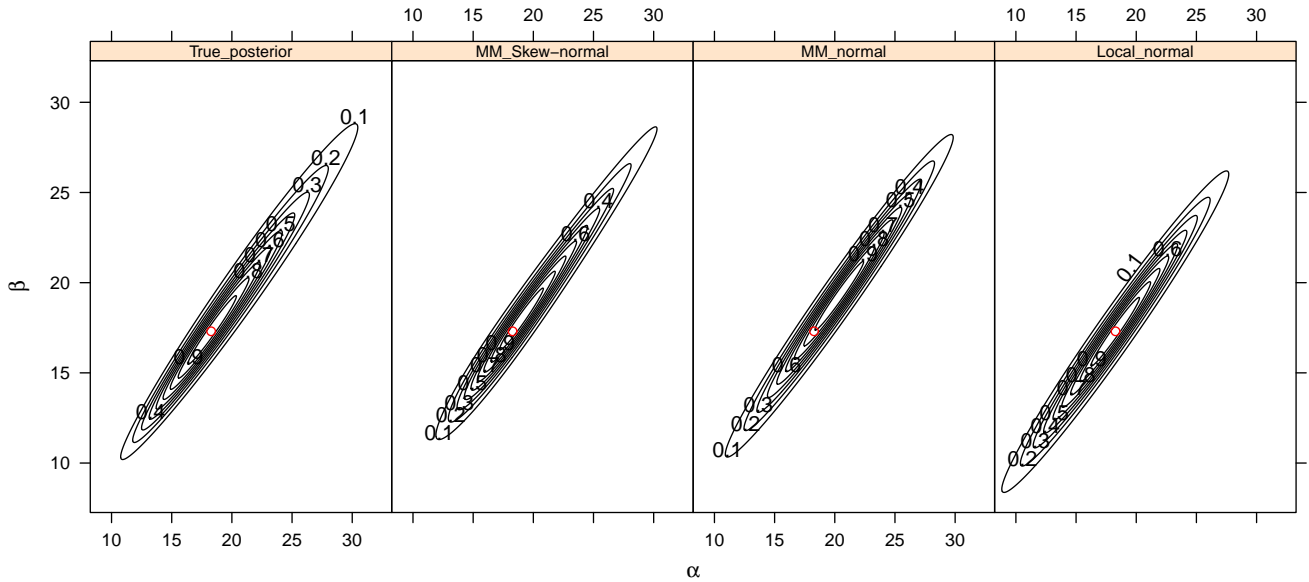}
\caption{Comparison between the true posterior density and approximate densities. The red point in each panel is the mode of the true posterior distribution. }
\label{fig:poll_contour}
\end{figure}

Figure \ref{fig:poll_contour} compares the posterior distributions of the hyper-parameters $(\alpha, \beta)$ and its approximate densities. The MM skew-normal approximation can capture the skewness of the posterior distribution while the MM normal and Local normal cannot. The distances between the mode of the true posterior and the one for the MM skew-normal approximation, MM normal, and Local normal are 0.87, 2.91, and 0, respectively. 
\begin{figure}[h]
\centering
\includegraphics[width=4in, angle=-90]{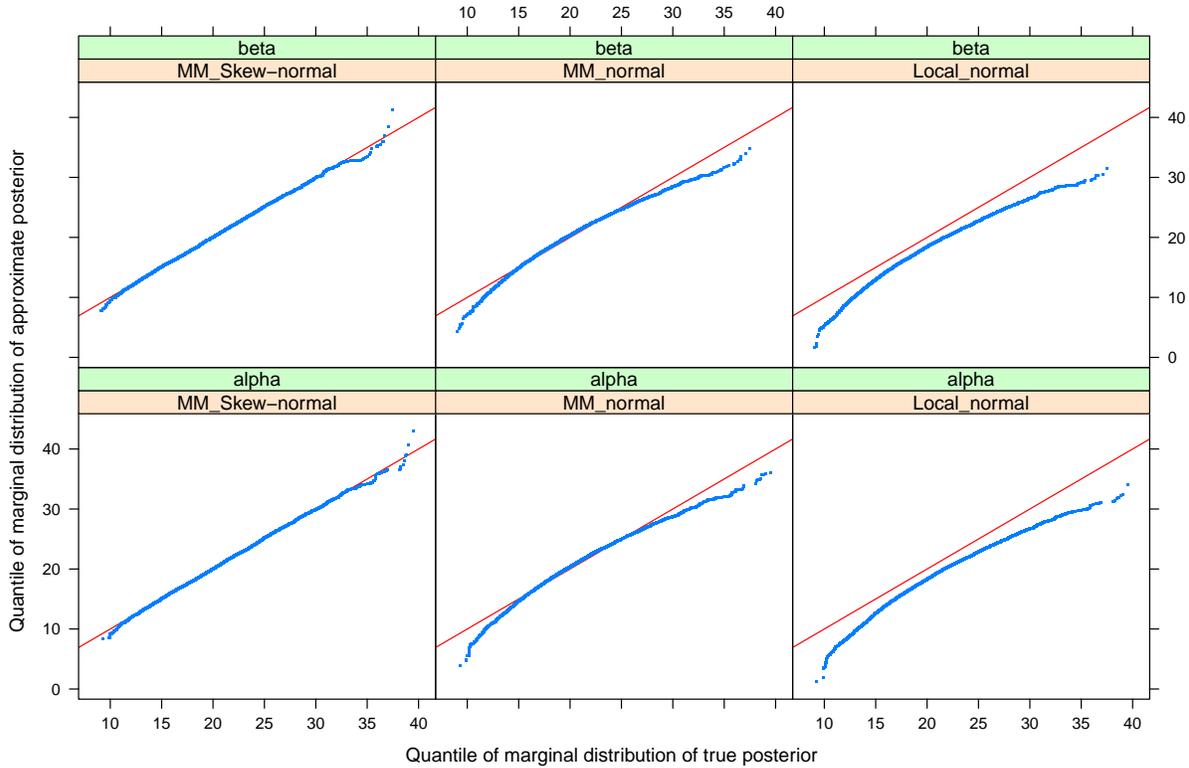}
\caption{Pair quantile comparisons among the true posterior density and its approximate densities. The red line is a 45-degree reference line in each panel.}
\label{fig:poll_pair}
\end{figure}

Besides the comparison of the joint density, the comparison of the marginal density is also of interest. Figure \ref{fig:poll_pair} is a plot of the quantiles of a marginal sample from the approximate densities against the quantiles of a marginal sample from the true posterior density with a sample size 10000. Panels in the first column are q-q plots of marginal densities of the MM skew-normal approximate density against the ones for the true posterior density. The second and third columns are for the MM normal approximate density and the Local normal approximate density against the true posterior density, respectively. Panels in the first row represent the marginal q-q plot for the parameter $\beta$ while the ones in the second row are for $\alpha$. If the two sets come from the same distribution, the points should fall approximately along the red reference line. Obviously, the MM skew-normal approximate density well approximates the true density while there is an unignorable departure from the MM normal approximation to the true density. The Local normal approximation is even worse.  

\begin{figure}[h]
\centering
\includegraphics[width=4in, angle=-90]{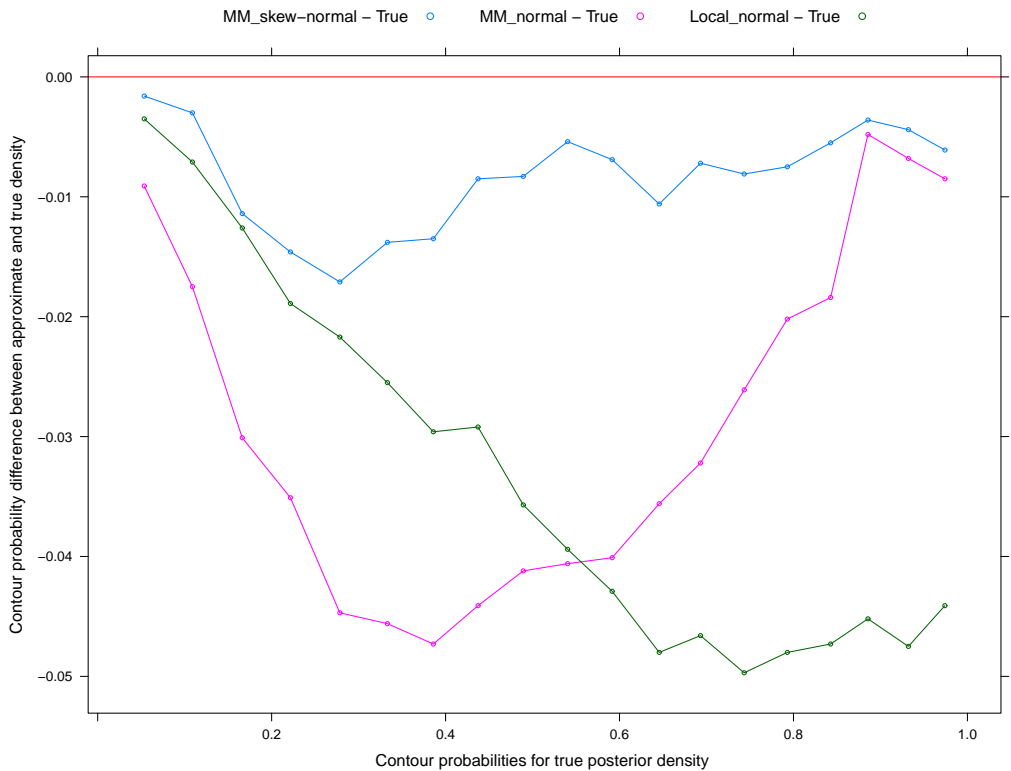}
\caption{Contour probability differences between approximate densities and the true posterior density under series of regions bounded by ellipsoids }
\label{fig:poll_prob}
\end{figure}

%
In order to have a deeper insight of the difference between the true posterior density and the approximation densities, we compute contour probabilities for three approximate density and true posterior function using CPA when $h_i'$s are chosen such that $T_i \in (0.05, 0.1, \cdots, 0.95)$. Contour probability differences between approximate densities and the true posterior density are plotted against the true contour probability. Figure \ref{fig:poll_prob} indicates that the MM skew-normal approximation method significantly outperforms the MM normal and the Local normal methods.

\subsection{Simulated Experiments}
In this section, the goal is to see how the likelihood modeling algorithm compares to a single machine algorithm run on the same data. Thus the data will have to be small enough for a single machine run to be possible. To assess the performance of likelihood modeling on distributed data for the logistic regression, we set up the experiments as follows:
\begin{itemize}
\setlength\itemsep{0.2em}
\item run: the number of simulations
\item m: log2 of the number of subset observations
\item r: log2 of the number of subsets
\item p: the number of the covariate variables
\item Coefficient vector $\theta =(1,\cdots, 1) $
\item Design matrix X with each row $x_i \overset{iid} \sim N^p(0,1) $,
\item Response variable Y with the element $y_i \sim Bernoulli(1/(1+\exp^{(-x_i^T\theta)})) $
\end{itemize}
\begin{figure}[h]
\centering
\includegraphics[width=4.5in, angle=-90]{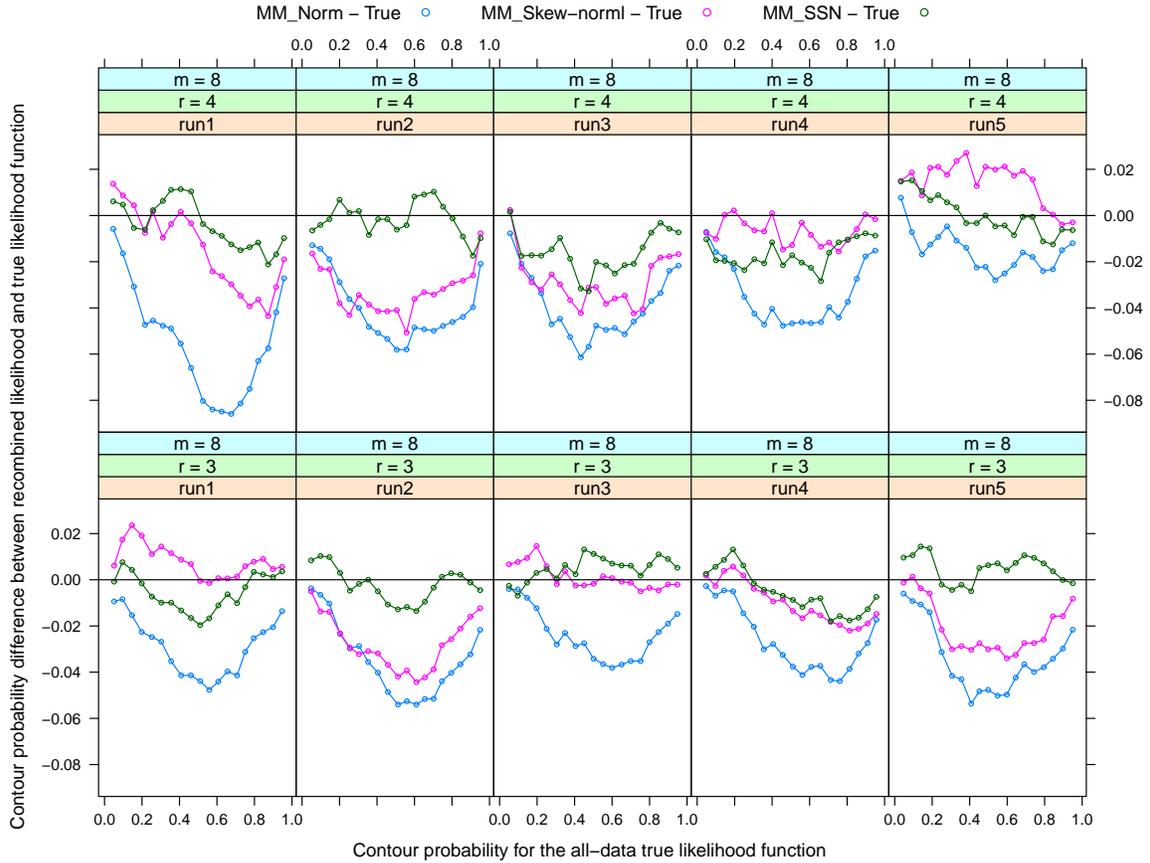}
\caption{Scatter plots of the contour probability differences between approximate likelihoods and the true likelihood, against the true contour probability in the cases of $m = 8$, $r = 3, 4$,$run = c(1, 2, \cdots, 5)$, and $\theta = (1, 1, 1, 1, 1)$}
\label{fig:prob.snmean}
\end{figure}
For each combination of $(m, r, run)$, the true likelihood function can be computed when data are generated with $p=5$ and stored in a single machine. In contrast, the MM skew-normal approximate likelihood, MM simplified skew-normal likelihood $(\text{MM}\_\text{SSN})$, and MM normal likelihood are estimated using the likelihood modeling algorithm when the same data are stored in a distributed cluster. Then, contour probabilities for both approximate likelihoods and true likelihood are estimated using the CPA. Figure \ref{fig:prob.snmean} displays plots of the contour probability differences against the true contour probability for several simulated cases. It is straightforward that the smaller the absolute contour probability difference is, the closer to the true likelihood function the approximate likelihood function is. The contour probabilities of the true likelihood range from 0.05 to 0.95 with a step size 0.05. Based on all panels, we can make a conclusion that the SN family are preferable to the normal family. And the MM simplified skew-normal model can be a good alternative candidate to replace the MM skew-normal model when we want to reduce computation workload for a large r.

\subsection{Computation Performance}
%

%

\begin{table}[!htb]
   
    \begin{subtable}{.4\linewidth}
      \centering
       \caption{}
\begin{tabular}{|p{4.2cm}|p{0.8cm}|p{0.8cm}|p{0.8cm}|  }
\hline
 & \multicolumn{3}{|c|}{Number of Nodes} \\ 
\hline
Methods & 10 & 50  &500 \\
\hline
Multi-machine MCMC &164.2  & 5 & 2.75 \\
\hline
Likelihood Modeling &2.04 &  &  \\
\hline
\end{tabular}
    \end{subtable}%
    \begin{subtable}{.6\linewidth}
      \centering
       \caption{}
\begin{tabular}{|p{0.6cm}|p{1.7cm}|p{1.7cm}|p{1.9cm}|  }
\hline
 & \multicolumn{3}{|c|}{r} \\ 
\hline
m &  8 & 11  &14 \\
\hline
 8&  126(3.96)& 128(7.81) & 661(7.90) \\
\hline
10 &534(6.1) &546(4.3)  &2598(6.01)  \\
\hline
12 &2104(52.1) &2165(107)  &10210(279)  \\
\hline
\end{tabular}
    \end{subtable} 
     \caption{Computation Performance. a) Running time (in hours) of the naive MCMC algorithm and likelihood modeling algorithm on clusters of different number of nodes for the case  p = 8, $2^r$ = 600,000, m = 7, iterations = 10,000. b) Running time (in seconds) on different size of data using likelihood modeling on the cluster of 10 nodes.}
     \label{diff_alg}
\end{table}
Scott 2013 \cite{Scott.2013} presents timings from a multi-machine MCMC algorithm for a single layer hierarchical logistic regression model on a 500-machine cluster and a 50-machine cluster. The running time to complete the job on a cluster of 500 machines and 50 machines is 2.75 hours and 5 hours, respectively. Scott concludes that a ten-fold reduction in computing resources only
produced a two-fold increase in compute time. In contrast, we run similar simulation experiments on a cluster of 10 machines using the likelihood modeling algorithm and MCMC algorithm (see Table \ref{diff_alg} (a)). All experiments are implemented on the WSC Cluster which consists of 10 nodes with total 200 cores, 128 GB RAM, 128.9 TB disk and 10 Gbps Ethernet interconnect. And all machines are running R version 3.3.1, Java 1.7.0$\_$07$\-$b10, Cloudera Hadoop 0.20.2$\-$cdh3u5 and Rhipe 0.75 \cite{Guha.2012}. The likelihood modeling algorithm reduced computation time in 80 folds with the same cluster setting. There might be a smarter way of setting up MCMC algorithm to reduce computation time. The bottleneck of the multi-machine MCMC algorithm is that the iterative algorithm is implemented as a chain of jobs where the output from each job is used as input to the next job.

The next test case is to run experiments to assess computation performance of the likelihood modeling algorithm. The test cases are all combinations of $r = (8, 11, 14), m = c(8, 10, 12)$ for run = 3, p = 10.  The value in each cell at Table \ref{diff_alg} (b) is the average of three runs while the value in parenthesis is the corresponding standard deviation of the three runs. It is noticing that the running time does not increase much when r increases from 8 to 11 with m fixed. Given m, the running time for r=14 is around 5 times the one for r = 11. The one possible explanation is that jobs for r = 11 make full use of containers which are idle when running jobs for r = 8.

\section{Discussion}
We have proposed an innovative divide \& recombine procedure to model the likelihood of generalized linear regression models on distributed datasets. There are many candidate models for likelihoods, just as there are many models for DM. Normal family and skew-normal family have been investigated to illustrate the likelihood modeling procedure. Also, we discussed two methods to estimate parameters of the given likelihood model family: MM with MCMC draws and Local method. Moreover, the contour probability algorithm was introduced to measure the similarity between approximate multivariate likelihood function and the true multivariate likelihood function. In terms of accuracy, the MM skew-normal likelihood model outperforms normal likelihood model in the application of CPA on Exit Poll data. On the computation point of view, the likelihood modeling definitely speeds up computation for generalized linear models, keeping the inference capability for big data. As the likelihood modeling procedure is designed to work in the divide \& recombine framework. In summary, the likelihood modeling algorithm can provide a relatively accurate estimate of the MLE of the parameters in the generalized linear model; it is well aligned with modern parallel and distributed computing architectures and is scalable to very large datasets.

Many approaches have been proposed to address the big data challenges. In the subsampling paradigm, there are the bags of little bootstrap (BLB) approach (Kleiner et al. \cite{Kleiner2014}), leveraging method (Ma et al. \cite{Ma2015}), resampling-based stochastic approximation method (Liang et al. \cite{Liang2013}). Other computationally efficient methods to draw approximate posterior samples (\cite{srivastava2015wasp} \cite{welling2011bayesian} \cite{ahn2012bayesian}\cite{broderick2013streaming} \cite{hoffman2013stochastic} \cite{Scott.2013} \cite{welling2011bayesian} \cite{neiswanger2013} \cite{wang2011} \cite{agarwal2011}\cite{wu2017average}). Lin et al. \cite{lin.2014} considered a distributed version of the trust region Newton method (TRON) to solve logistic regression and linear support vector machine (SVM) in Spark.

 Nevertheless, the likelihood modeling has some limitations. First of all, LM is constructed under the assumption that all observations are independent. Second, MCMC sampling method is used to generate a sample based on the subset likelihood function. There is a trade-off between computation time and the effective sample, especially in high dimension space. There are two possible future work. One of the potential future works is to modify methods within the $D\&R$ framework for non-iid data. Another follow-up work is to investigate more efficient strategies to capture information of the subset likelihood.

\bibliographystyle{unsrt}
\bibliography{reference}
\section{Appendix A -- Skew-normal}

\subsection{Univariate case}
To illustrate how to estimate parameters of the skew-normal, we introduce some basic definitions and relevant properties of the skew-normal family (Azzalini and Valle \cite{Azzalini.2008}). 
The skew-normal density function, in one-dimensional case, is given by
$$
f_1(\theta|\xi,\omega^2,\alpha) = \frac {2}{\sqrt{2 \pi\omega^2}} \exp \left( -\frac{ (\theta - \xi)^2}{2\omega^2}  \right)\Phi(\alpha(\frac{\theta-\xi}{\omega})), \, \xi, \alpha \in \mathbb{R}, \omega \in \mathbb{R}^{+},
$$
where $\Phi$ denotes the cumulative distribution function (CDF) of the standard normal distribution; $\xi, \omega$, and $\alpha$ are the location, scale, and shape parameters, respectively. We say $\Theta \sim SN(\xi,\omega^2, \alpha)$ if random variable $\Theta$ has density function $f_1(\theta|\xi,\omega^2,\alpha)$.

Suppose $\Theta \sim SN(\xi,\omega^2, \alpha)$ and $\Theta = \xi + \omega Z $, then
$$
Z = (\Theta - \xi)/\omega,
$$
which is the "normalized" random variable with a distribution $SN(0,1,\alpha)$. It's worth noting that Z has non-zero mean if $\alpha \neq 0$. More specifically, the mean, variance, and skewness of Z are
$$
\mu_Z= b\delta, \quad \sigma^2_Z = 1 - \mu_Z^2, \quad \gamma_Z = \frac{4-\pi}{2}\frac{\mu_Z^3}{(1 - \mu_Z^2)^{3/2}},
$$
where $b = \sqrt{2/\pi}$ and $\delta = \alpha/\sqrt{(1+\alpha^2)}$. Therefore, the mean, variance and skewness of $\Theta$ are
\begin{align}
\mu_{\Theta} &= E[\Theta] = \xi + \omega \mu_Z,\\
\sigma_\Theta^2 &= var[\Theta] = \omega^2(1 - \mu_Z^2),\\
\gamma_\Theta &= E\left[(\frac{\Theta - \mu_\Theta}{\sigma_\Theta})^3 \right] =  \frac{4-\pi}{2}\frac{\mu_Z^3}{(1 - \mu_Z^2)^{3/2}},
\end{align}
which form the centered parametrization of $SN(\xi,\omega^2, \alpha)$. Also
these three equations imply the way to estimate parameters of $SN(\xi,\omega,\alpha)$. Given a random sample $\theta_1, \theta_2, \cdots, \theta_n$ from distribution $SN(\xi,\omega,\alpha)$, we can calculate sample mean $\hat{\mu_\Theta}$, sample variance $\hat{\sigma_\Theta^2}$ and sample skewness $\hat{\gamma_\Theta}$. By solving equations (3), (2), (1), sequentially, we obtain
\begin{align}
\hat{\mu}_Z &= \frac{\hat{c}}{\sqrt{1+\hat{c}^2}},\\
\hat{\alpha} &= \frac{\hat{\mu}_Z}{\sqrt{b^2 - \hat{\mu}_Z^2}},\\
\hat{\omega}^2 &=\frac{ \hat{\sigma}_\Theta^2}{1- \hat{\mu}_Z^2},\\
\hat{\xi} &= \hat{\mu}_\Theta - \hat{\omega} \hat{\mu}_Z,
\end{align}
where $\hat{c} = (\frac{2\hat{\gamma}_\Theta}{4 - \pi})^{1/3}$.

The parameters estimation is straightforward when the sample is available. However, not all sample can successfully derive estimates of the parameters. As a matter of fact, $$\delta \in (-1, 1) \Longrightarrow \mu_Z \in (-b, b).$$
Therefore, 
$$\gamma_\Theta \in  (-\frac{4-\pi}{2}\frac{b^3}{(1 - b^2)^{3/2}}, \frac{4-\pi}{2}\frac{b^3}{(1 - b^2)^{3/2}}) \approx (-0.9952717, 0.9952717).$$
If $\hat{\gamma}_\Theta$ derived from the sample falls in above region, then we call $(\hat{\mu}_\Theta,\hat{\sigma}^2_\Theta,\hat{\gamma}_\Theta)$ admissible; otherwise inadmissible. As the normal density function is a special case of the skew-normal density function with $\alpha = 0$. If a normal density is considered as a candidate approximate function for the logistic likelihood function, then the parameters of the normal density can be easily estimated by the sample mean and the sample standard error.

\subsection{Multivariate case}
The Multivariate SN distribution has been widely discussed by Azzalini, Dalla Valle and Capitanio. Similar to the univariate case, the p-dimensional SN density function is defined by
$$
f_p(\theta|\xi,\Omega,\alpha) =\frac {2}{\sqrt{(2 \pi)^p |\Omega|}} \exp \left( -\frac{1}{2} (\theta - \xi)^\intercal \Omega^{-1} (\theta - \xi) \right)\Phi(\alpha^\intercal \omega^{-1}(\theta -\xi)), \, \xi, \alpha \in \mathbb{R}^p, \Omega \in \mathbb{R}^{p\times p},
$$
where $\Omega$ is a $p\times p$ positive definite matrix, $\xi$ is a vector location parameter, $ \alpha$ is a vector shape parameter, and $\omega$ is a diagonal matrix formed by the square root of the diagonal of $\Omega$. We say $\Theta \sim SN(\xi,\Omega, \alpha)$ if a multivariate random variable $\Theta$ has density function $f_p(\theta|\xi,\Omega,\alpha)$.

To derive the estimating formulas, let $\Theta = \xi + \omega Z$. Then
$$
Z = \omega^{-1}(\Theta - \xi),
$$
which is the 'normalized' variable with distribution $SN(0,\overline{\Omega},\alpha )$, where $\overline{\Omega} = \omega^{-1}\Omega \omega^{-1}$. It is worth noting that the diagonal elements of $\overline{\Omega}$ are all ones. Let $b = \sqrt{2/\pi}$, $\delta = (1 + \alpha^\intercal \overline{\Omega} \alpha)^{-1/2} \overline{\Omega}\alpha$ and $\gamma_{zi} =  \frac{4-\pi}{2}\frac{\mu_{zi}^3}{(1 - \mu_{zi}^2)^{3/2}}$, then 
$$
\mu_Z = E[Z] = b\delta,\quad  \Sigma_Z = var[Z] = \overline{\Omega} - \mu_Z \mu_Z^\intercal, \quad \gamma_Z = (\gamma_{z1}, \dots, \gamma_{zp}).
$$
Therefore, it is trivial that
\begin{align*}
\mu_\Theta &= E[\Theta] = \xi + \omega \mu_Z,\\
\Sigma_\Theta &= var[\Theta] = \omega \Sigma_Z \omega = \Omega - \omega\mu_Z \mu_Z^\intercal\omega,\\
\gamma_\Theta &= \gamma_Z.
\end{align*}

The derivation of the parameters estimation for the multivariate skew-normal density is similar to univariate case. To simplify the notation, let $\sigma_Z = \sqrt{diag(\Sigma_Z)}$ and $\sigma_\Theta = \sqrt{diag(\Sigma_\Theta)},$  i.e. the square root of the diagonal of the variance matrix of Z and $\Theta$, respectively.
Given a multivariate random variable sample $\theta_1, \dots, \theta_n$ drawn from distribution $SN(\xi, \Omega, \alpha)$, sample mean $\hat{\mu}_\Theta$, sample covariance $\hat{\Sigma}_\Theta$, and componentwise skewness $\hat{\gamma}_{\Theta}$ can be easily computed. Then $\hat{\mu}_Z$ can be obtained by using (4). Therefore, the parameters will be estimated as follows:
\begin{align}
\hat{\delta}  &= \hat{\mu}_Z/b, \quad \hat{\sigma}_Z =\sqrt{diag(I - \hat{\mu}_Z \hat{\mu}_Z^\intercal)},\\
\hat{\omega} &= diag(\hat{\sigma}_Z^{-1}\hat{\sigma}_\Theta), \quad \hat{\xi} = \hat{\mu}_\Theta - \omega \hat{\mu}_Z,\\
\hat{\Omega} &= \hat{\Sigma}_\Theta + \hat{\omega}\hat{\mu}_Z \hat{\mu}_Z^\intercal\hat{\omega}, \quad \hat{\alpha} = \frac{\hat{\overline{\Omega}}^{-1}\hat{\delta}}{\sqrt{1- \hat{\delta}^\intercal\hat{\overline{\Omega}}^{-1}\hat{\delta}}},
\end{align}
where $diag(\hat{\sigma}_Z^{-1}\hat{\sigma}_\Theta)$ is a main diagnal matrix with components $\hat{\sigma}_{Zi}^{-1}\hat{\sigma}_{\Theta i},\quad i = 1,\cdots, p$.

There several properties of this estimation method. First of all, this method enables us to estimate parameters of the multivariate skew normal in a closed form, rather than in an iterative approach, which greatly reduces the computational cost. The estimation procedure for the multivariate case is an extended version of the univariate case since the multivariate case reduces to the univariate case when p = 1.  Given $(\xi, \Omega, \alpha)$, there must exist only one corresponding $(\mu, \Sigma,\gamma)$. However, not vice versa. As a matter of fact, the corresponding $(\xi, \Omega, \alpha)$ may not exist even though $(\mu, \Sigma,\gamma)$ satisfy the constraint that $\Sigma$ is positive definite. Additional constraints should include
$$ \gamma_{\Theta i} \in (-\frac{4-\pi}{2}\frac{b^3}{(1 - b^2)^{3/2}}, \frac{4-\pi}{2}\frac{b^3}{(1 - b^2)^{3/2}}) \approx (-0.9952717, 0.9952717), \quad i = 1, \cdots, p, $$
$$1- \delta^\intercal\overline{\Omega}^{-1}\delta > 0.$$
For the first constraint, it is implicit in the genesis of the multivariate skew-normal random variable. Because the marginal distribution of a subset of the components of the multivariate skew normal random variable is still a skew-normal random variable (Azzalini \& Dalla Valle \cite{Azzalini.1996}). For the second constraint, it is straightforward. In order to obtain the parameters estimates, we resample the data until $(\xi, \Omega, \alpha)$ can be estimated. Recall that we assume the sample of the logistic likelihood function is a good approximate sample of the SN distribution. Simulation studies show that $(\xi, \Omega, \alpha)$ usually can be successfully estimated with a sample drawn from the subset logistic likelihood for the first time when the subset likelihood function is not too flat. If the number of observations in a subset is small, the corresponding likelihood is flat in the neighborhood of the MLE. Therefore, the skewness of a sample drawn from a flat density function is very sensitive to the sample.

\section{Appendix B -- Concavity}
To prove that the multivariate skew-normal density is concave, we assume $\theta \sim SN(\xi,\Omega, \alpha)$. Then the log density function is
$$ \log f(\theta) = -\frac{1}{2} \log \left( \frac{1}{4} (2\pi)^p |\Omega| \right) -\frac{1}{2} (\theta - \xi)^\intercal \Omega^{-1} (\theta - \xi)  + \log \Phi\left( \lambda^\intercal (\theta - \xi) \right), $$

where $\lambda^\intercal = \alpha^\intercal \omega^{-1} $. The first and second order relevant derivatives respect to $\theta$ are

$$\frac {\partial} {\partial \theta_k} \log f(\theta) =  - (\theta - \xi)^\intercal \Omega_{\cdot k}^{-1} + \frac {\lambda_k \phi(\lambda^\intercal (\theta - \xi))}{\Phi(\lambda^\intercal (\theta - \xi))},$$

$$H_{j,k} = \frac {\partial^2} {\partial \theta_j \theta_k} \log f(\theta) =  -\Omega_{jk}^{-1} + \lambda_j \lambda_k \frac { \phi'(\lambda^\intercal (\theta - \xi)) \Phi(\lambda^\intercal (\theta - \xi)) - \phi^2(\lambda^\intercal (\theta - \xi))} {\Phi^2(\lambda^\intercal (\theta - \xi)) },$$

Where 
$$ \phi(\lambda^\intercal (\theta - \xi)) = \frac {1} {\sqrt{2 \pi}} e^{ - \frac{1}{2} (\lambda^\intercal (\theta - \xi))^2 }, $$

$$ \Phi (\lambda^\intercal (\theta - \xi)) = \int_{- \infty}^{ \lambda^\intercal (\theta - \xi)} \frac {1} {\sqrt{2 \pi}} e^{ - \frac{1}{2} x^2 } dx, $$

$$ \phi'(\lambda^\intercal (\theta - \xi)) = \frac {-1} {\sqrt{2 \pi}} e^{ - \frac{1}{2} (\lambda^\intercal (\theta - \xi))^2 } \lambda^\intercal (\theta - \xi). $$

The $\log f(\theta)$ is concave if and only if Hessian matrix H is negative semidefinite. Let 
$$g(t) =  \frac{\phi'(t) \Phi(t) - \phi^2(t)}{\Phi^2(t)} = - \frac{\phi(t)(t \Phi(t) + \phi(t))}{\Phi^2(t)}.$$
It is trivial to prove that $t\Phi(t) + \phi(t) \geq 0, t \in \mathbb{R}$. Therefore, it is straightforward that $g(t) \leq 0, t \in \mathbb{R}$ and 
$$ v^T H v = - v^T \Omega^{-1} v + g(\lambda^\intercal (\theta - \xi))(\lambda^\intercal v)^2 < 0, v \in \mathbb{R}^p/\{0\}.$$
\section{Appendix C -- Poll Exit}

\begin{table}[]
\centering
\caption{California Democratic Poll Exit}
\label{polldata}
\begin{tabular}{llllllll}
fips & total\_voters & sample\_voters & sample\_clinton & fips & total\_voters & sample\_voters & sample\_clinton \\
6001 & 199445        & 100            & 52              & 6059 & 226598        & 165            & 93              \\
6003 & 241           & 198            & 94              & 6061 & 30402         & 112            & 69              \\
6005 & 3769          & 150            & 75              & 6063 & 2747          & 173            & 65              \\
6007 & 24202         & 103            & 33              & 6065 & 123078        & 152            & 90              \\
6009 & 5126          & 104            & 54              & 6067 & 119943        & 166            & 88              \\
6011 & 1275          & 100            & 45              & 6069 & 3504          & 101            & 62              \\
6013 & 117523        & 122            & 68              & 6071 & 124555        & 124            & 69              \\
6015 & 2388          & 179            & 81              & 6073 & 253744        & 138            & 75              \\
6017 & 20130         & 166            & 79              & 6075 & 153003        & 140            & 83              \\
6019 & 55285         & 155            & 92              & 6077 & 42003         & 121            & 81              \\
6021 & 1321          & 177            & 95              & 6079 & 33266         & 175            & 99              \\
6023 & 19470         & 153            & 46              & 6081 & 77763         & 189            & 118             \\
6025 & 8597          & 196            & 129             & 6083 & 46898         & 184            & 97              \\
6027 & 1749          & 124            & 53              & 6085 & 181757        & 162            & 105             \\
6029 & 33340         & 112            & 60              & 6087 & 45486         & 150            & 59              \\
6031 & 6623          & 163            & 98              & 6089 & 12290         & 113            & 58              \\
6033 & 5189          & 127            & 62              & 6091 & 493           & 183            & 81              \\
6035 & 1516          & 198            & 91              & 6093 & 3962          & 106            & 39              \\
6037 & 1035968       & 144            & 61              & 6095 & 55903         & 177            & 106             \\
6039 & 8688          & 101            & 54              & 6097 & 88257         & 128            & 70              \\
6041 & 47288         & 123            & 71              & 6099 & 27885         & 117            & 69              \\
6043 & 2048          & 115            & 62              & 6101 & 4340          & 120            & 65              \\
6045 & 7390          & 140            & 43              & 6103 & 3117          & 154            & 86              \\
6047 & 12577         & 126            & 61              & 6105 & 1568          & 103            & 40              \\
6049 & 551           & 200            & 81              & 6107 & 14414         & 168            & 106             \\
6051 & 1681          & 118            & 61              & 6109 & 5557          & 182            & 100             \\
6053 & 30311         & 146            & 90              & 6111 & 85219         & 130            & 65              \\
6055 & 12242         & 177            & 99              & 6113 & 24260         & 163            & 81              \\
6057 & 14154         & 187            & 75              & 6115 & 3387          & 196            & 85             
\end{tabular}
\end{table}

\end{document}